\documentclass[journal=jacsat,manuscript=article]{achemso}
\usepackage[version=3]{mhchem} 
\usepackage{graphicx}
\usepackage{dcolumn}
\usepackage{bm}
\usepackage{mathtools}
\usepackage{amsmath}
\usepackage{amsfonts}
\usepackage{amssymb}
\usepackage{enumerate}
\usepackage{mathtools}
\usepackage{float}
\usepackage{xfrac}
\usepackage{hyperref}
\usepackage{tabularx}
\usepackage{xcolor}
\hypersetup{
    colorlinks=true, 
    linktoc=all,     
    linkcolor=black,  
    citecolor=black,
}

\author{Rahul Chand}
\altaffiliation{These authors contributed equally to this work}
\affiliation{Department of Physics, Indian Institute of Science Education and Research (IISER), Pune 411008, India}

\author{Chaudhary Eksha Rani}
\altaffiliation{These authors contributed equally to this work}
\affiliation{Department of Physics, Indian Institute of Science Education and Research (IISER), Pune 411008, India}

\author{Diptabrata Paul}
\affiliation{Department of Physics, Indian Institute of Science Education and Research (IISER), Pune 411008, India}
\alsoaffiliation{Present address: Faculty for Physics and Earth Sciences, Leipzig University}

\author{G V Pavan Kumar}
\email{pavan@iiserpune.ac.in}
\affiliation{Department of Physics, Indian Institute of Science Education and Research (IISER), Pune 411008, India}

\title{Emergence Of Directional Rotation In Optothermally Activated Colloidal System}

\begin{document}

\begin{abstract}
We experimentally demonstrate the emergence of directional rotation in thermally active-passive colloidal structures under optical confinement. The observed handedness of rotation of the structure can be controlled by changing the relative position of the constituent colloids. We show that the angular velocity of rotation is sensitive to the intensity of the incident optical fields and the size of the constituent colloidal entities. The emergence of rotational dynamics can be understood in the context of asymmetric temperature distribution in the system and the relative location of the active colloid, which creates a local imbalance of optothermal torques to the confined system. Our work demonstrates how localized optothermal fields lead to directional rotational dynamics without explicitly utilizing spin or orbital angular momentum of light. We envisage that our results will have implications in realizing Brownian engines, and can directly relate to rotational dynamics in biological and ecological systems.
\end{abstract}

\section{Introduction}
Systems which are out-of-equilibrium are ubiquitous in nature {\cite{collective_motion,hydrodynamics_of_softmatter, Living_matter_phy_reports,broken_detailed_balance_out-of-equilibrium,bacterial_active_matter,odd_dynamics_chiral_crystal,collective_behaviour_active_matter}}. These systems exhibit interesting emergent dynamics such as dynamic self-assembly \cite{whitesides2002self, copeland2009bacterial, vincenti2019magnetotactic}, pattern formation \cite{ bacteria_pattern_formation, bactterial_pattern_minimal_mechanism, steager2008dynamics, curatolo2020cooperative} and collective motility (both in terms of translation and rotation) \cite{collective_motion, odd_dynamics_chiral_crystal, bacterial_colony_rotation_vicsek_1996, magnetotactic_bacteria_rotating_mag_mield, individual_to_collective_motion_zhang2012, collective_motion_Ecoli, Goldstein_bacteria_directed_motion, bacteria_moving_slowly_2021}. These phenomena emerge due to the mutual interaction between the individual elements of these collective systems. Understanding the underlying principle behind these collective phenomena motivates researchers to emulate similar behavior in laboratory environments, which is one of the key goals of active matter research {\cite{spontaneous_helically_assembled_active_matter,thermodynamics_active_matter, How_far_equilibrium_active_matter,hydrodynamic_of_active_matter,computational_model_active_matter}}. In this regard, artificial soft matter systems such as synthetic colloids can be used as test beds. Using such systems, we can study the interaction between individual entities that lead to collective behaviour, and understand the role of environmental cues that drive such systems.

In this context, two interrelated tasks have been pursued: first, a variety of colloidal matter has been realized and studied in the context of active systems {\cite{active_colloidal_rollar, self-propelled_colloidal_motors, active_particles_crowded_evironment, light-activated_colloidal_active_matter, volpe_active_colloidal_molecules, practical_guide_active_colloids}}. Second, diverse methods including chemical {\cite{cell_stimulation_chemical_gradients,self-propulsion_micromotors_cleaning, beyond_platinum_2014, Au@TiO2_active_matter_Juliane_Simmchen}}, electric {\cite{on-chip_electric_manipulation, solodkov2020electrically, electrically_powered_locomotion}}, magnetic \cite{magnetic_rotor_Golestanian,magnetically_actuated_golestanian,magnetic_micromachines_cargo_transport,micro_and_nanobots_oscillating_magtic_field, zhou2021magnetically, tierno2021transport}, and optical fields {\cite{ashkin1986observation, revolution_in_optical_manipulation, optical_manipulation_Dholakia, optical_manipulation_micro-to-macro, marago_2018_optical_tweezers, light_driven_plasmonic_motor_2018, nanoscale_inorganic_2019, light_induced_rotation_around_optical_fiber, optical_rotation_and_thermometry, initiating_revolution_2021_Dholakia, nan2022creating, ding2022programmable}} have been innovated that can act as an environmental cue to translate and rotate soft systems. 

The point of interest to the current study is the rotational dynamics. The emergence of rotational degrees of freedom has been observed in a variety of natural systems \cite{bacterial_colony_rotation_vicsek_1996, magnetotactic_bacteria_rotating_mag_mield, Goldstein_bacteria_directed_motion, vincenti2019magnetotactic, spermatozoon_rotation_1992, rotating_bacteria_without_tethering, chemotactic_bacteria_2019}, and emulating them using artificial soft matter systems \cite{active_colloidal_rollar, microscopic_engine_schmidt2018, peng2020opto, janus_microengine_Marago} is an interesting but challenging task.

Among the available tools, optical fields have emerged as versatile and complementary methods to not only drive soft systems but also confine them {\cite{trapping_material_world_Dholakia,information_flow_cichos,optical_trapping_with_structure,volpe_roadmap_optical_tweezer}}. In the context of generating rotational degrees of freedom, the spin and the orbital angular momentum of light {\cite{torque_volpe, zhao_direct_angular_momentum,rotation_by_elliptical_polarized, dholakia_2016_lg_beam_rotation,nanoparticle_rotation_lg_beam,passive_orbital_hydrodynamics, rotational_simulation_Ishihara}} has been extensively utilized. A relevant question to ask is: can we introduce rotational degrees of freedom in an active colloidal system without using the spin or orbital angular momenta of light? This will create new and complementary opportunities to manipulate soft systems without having to depend on the angular momentum of light.   

One of the ways to do so is to utilize optothermal effects {\cite{laser_induced_heating}}, where localized optical heating can create asymmetric temperature gradients, thereby breaking the symmetry that results in rotation {\cite{thermophoretic_microgear, microscopic_engine_schmidt2018, optothermal_rotation, peng2020opto, ding2022universal, janus_microengine_Marago}}. To gain control over this approach, there is now an imperative to identify the material parameters of thermally active colloids and the nature of optothermal fields that generate rotation. Specifically, generating directional rotation in the artificial active matter has emerged as an important task {\cite{Au@TiO2_active_matter_Juliane_Simmchen,volpe_active_colloidal_molecules,thermophoretic_microgear}}, which needs attention for both fundamental understanding and biological and technological applications.

\begin{figure}[t]
\centering
\includegraphics[width = 240 pt]{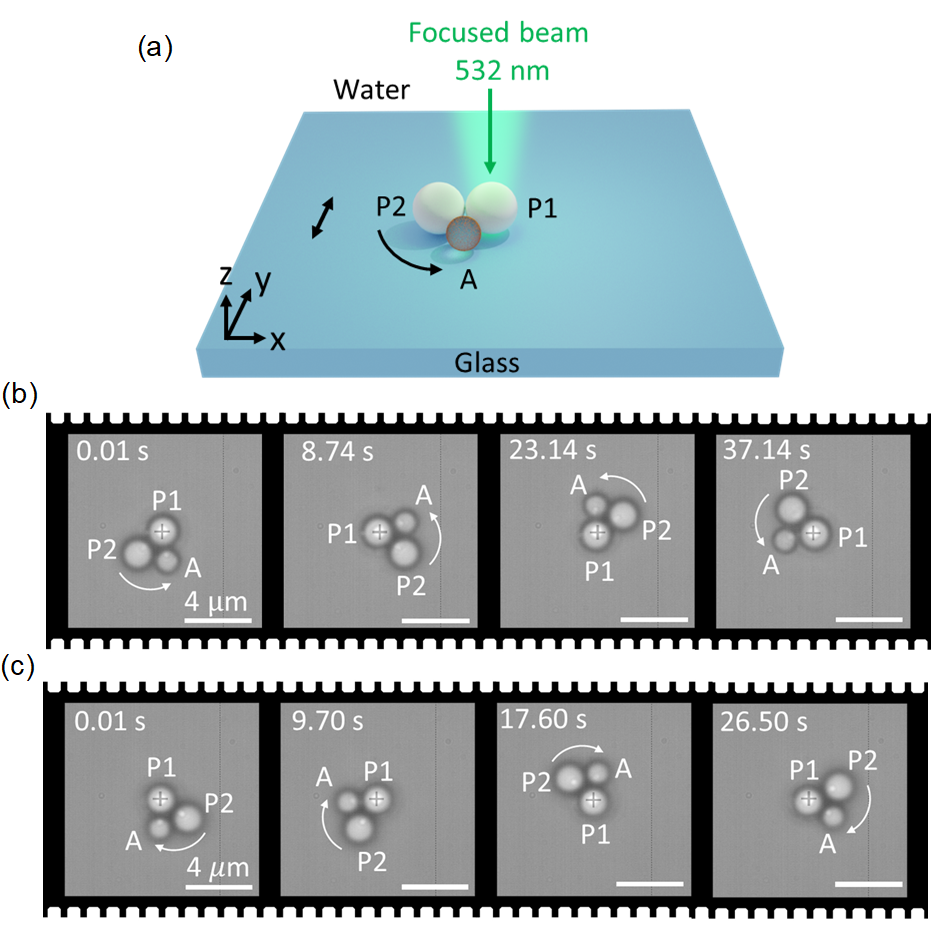}
\caption{Rotation of a trimer structure in an optical field. (a) Schematic of the experiment. In a focused linearly polarized Gaussian laser beam of wavelength 532 nm, a trimer structure with two passive melamine formaldehyde colloids (P1 and P2) of diameter 2.0 {\textmu}m and one thermally active polystyrene colloid (A) with diameter 1.3 {\textmu}m exhibits rotation. The beam is polarized along the $y$-axis as indicated by the double-sided black arrow. The direction of rotation is indicated by a curved black arrow. In (b) and (c) time series of rotation of two trimer structures with different handedness have been shown at 5.7 mW/\textmu$\textrm{m}^2$ laser intensity. The position of the beam center is indicated by the grey ‘+’ symbol.}
\label{fig:Figure_1}
\end{figure}

Motivated by this, here we present our experimental observation of generating directional rotation in colloidal structures composed of thermally active (A) and passive (P) colloids under 532 nm laser beam illumination. Figure \ref{fig:Figure_1}(a) shows a typical schematic of rotation in a trimer structure where P1, P2, and A are the passive and thermally active colloids respectively. Here the passive colloids P1 and P2 are almost identical, the only difference between them is their position with respect to the laser beam center. The handedness of this rotational motion depends on the relative position of the colloid A in the structure, which is evident from Figure \ref{fig:Figure_1}(b) and Figure \ref{fig:Figure_1}(c). Furthermore, we have shown how such rotation can be controlled by using a combination of colloids and the intensity of optical fields. We will first discuss the experimental methods used in the realization of such dynamics followed by a detailed discussion of the directional rotation. 

\section{Results and discussion}
In our experiments, we have used two types of colloids (i) passive colloids denoted by P and (ii) thermally active colloids denoted by A. Melamine formaldehyde (MF) (diameter 2.0 {\textmu}m) and polystyrene (PS) particles (diameter 1.3 {\textmu}m) have been used as passive colloids and iron oxide doped polystyrene (PS) particles (diameter 1.3 {\textmu}m) have been used as thermally active colloids (obtained from Microparticles GmbH). The iron-oxide nanoparticles in the thermally active colloids lead to the absorption of incident optical field and the consequent generation of thermal fields {\cite{optothermal_evolution_ACS}} (Supporting Information S1). These thermally active and passive colloids are dispersed in an aqueous suspension. 10 {\textmu}L of the sample solution is dropcasted onto a chamber of height $\simeq$ 100 {\textmu}m, enclosed between two glass coverslips. The sample cell is then placed in a dual-channel optical microscopy setup as shown in Supporting Information S2. A linearly polarized Gaussian laser beam of wavelength 532 nm is focused on the lower coverslip using a 100$\times$0.95 NA objective lens. The signal is then collected through a 100$\times$1.49 NA oil immersion objective lens and projected to a fast camera to record the dynamics of colloids at 100 fps. The experimental laser beam profile is shown in Supporting Information S3. The trajectory of individual colloids are extracted from the recorded videos \cite{trackmate_fiji}.

\subsection{Dynamics of colloidal structures under optothermal confinement} 

\begin{figure}[ht]
\includegraphics[width = 450 pt]{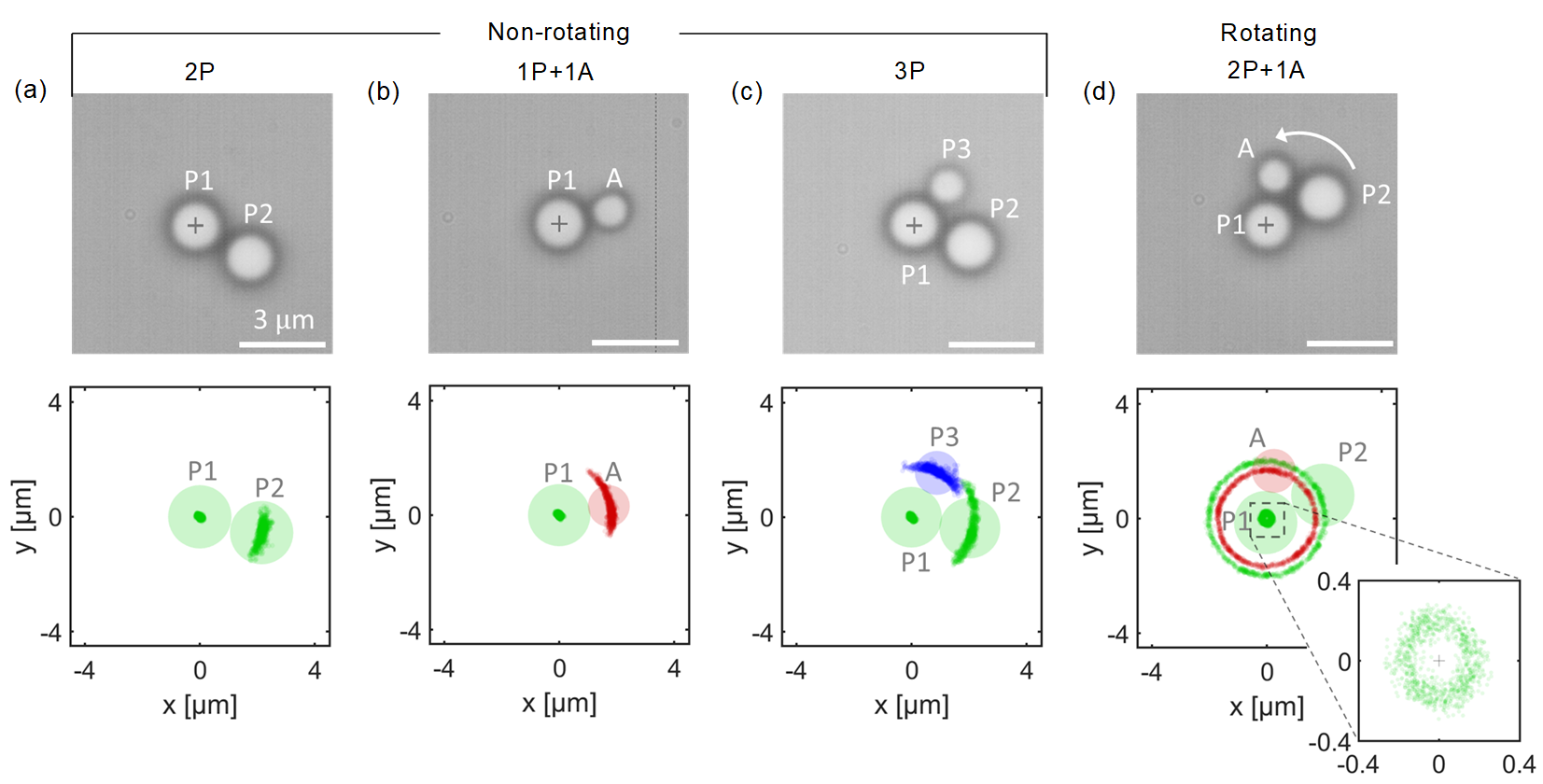}
\caption{Dynamics of structures composed of passive and thermally active colloids at $5.7$ mW/\textmu$\textrm{m}^2$ laser intensity. No rotational motion is observed in dimer structures composed of (a) two passive MF colloids (P1, P2) and (b) one passive MF colloid (P1) and one thermally active PS colloid (A). (c) Passive trimer structure composed of two passive MF colloids (P1, P2) and one passive PS colloid (P3) does not exhibit any rotational dynamics. (d) Active trimer composed of two passive MF colloids (P1, P2) and one thermally active PS colloid (A) exhibits counterclockwise rotation. Inset shows the magnified position distribution of the central passive colloid (P1) which is directly trapped at beam center. Here the diameters of the colloids are $\mathrm{d_{P1}}=\mathrm{d_{P2}}=2.0$ {\textmu}m and $\mathrm{d_{P3}}=\mathrm{d_{A}}=1.3$ {\textmu}m and all the position distributions are taken over 50 seconds.}
\label{fig:Figure_2}
\end{figure}

To understand the colloidal configuration leading to the emergence of rotation, we have systematically studied four different structures under optical confinement as shown in Figure \ref{fig:Figure_2}(a)-\ref{fig:Figure_2}(d). The structures are composed of (a) two passive colloids (P1, P2), (b) one passive (P1) and one thermally active colloid (A), (c) three passive colloids (P1, P2, P3), and (d) two passive (P1, P2) and one thermally active colloid (A) respectively. Out of these four structures, only the structure shown in Figure \ref{fig:Figure_2}(d) exhibits rotational motion. The colloids in the dimer structures, in Figure \ref{fig:Figure_2}(a) and Figure \ref{fig:Figure_2}(b), are confined radially to the beam center as a result of optical gradient forces and do not exhibit any rotational motion. If one more passive polystyrene colloid (P3) of diameter $\mathrm{d_{P3}}=1.3$ {\textmu}m is added to the structure shown in Figure \ref{fig:Figure_2}(a), a trimer structure is formed (Figure \ref{fig:Figure_2}(c)), aided by the optical gradient force-induced radial confinement. The non-rotating nature signifies the passivity of the system and therefore we indicate this as a passive trimer (PT) structure. But, when P3, in Figure \ref{fig:Figure_2}(c), is replaced by a thermally active PS colloid (A) of the same size, rotational motion is observed as shown in Figure \ref{fig:Figure_2}(d), thereby rendering an active trimer (AT) structure (see SI video 1). The position distribution of P1, shown in the inset, indicates the rotation of the whole structure about a common center. The angular velocity of such a rotating structure is obtained by calculating the differential cross-correlation function from the trajectory ($x,y$) of individual colloids as 
\begin{equation}
\begin{split}
    dccf(\delta t) &=\frac{\langle x(t)y(t+\delta t)\rangle-\langle y(t)x(t+\delta t)\rangle}{\sqrt {\langle x^2 (t)\rangle \langle y^2 (t)\rangle}},\\
    &= {A} e^{-\frac{\delta t}{\delta t_\textrm{ot}}}\sin(\omega \delta t),
\end{split}
\label{equation 1}
\end{equation}
 where $\delta t$, $\omega$ and $\delta t_{ot}$ are lagtime, the angular velocity of rotation, and the characteristic timescale of the trap respectively {\cite{microscopic_engine_schmidt2018}}. The angular velocity of rotation for AT, shown in Figure \ref{fig:Figure_2}(d), is obtained as $\omega=0.60$ rad/s. The magnitude of torque experienced by the system of colloids during this process can be estimated considering the moment of inertia ${I}_{i}$ of corresponding colloids about the center of rotation ({Supporting Information} S4) as 
 \begin{equation}
     |\tau|=\sum_i \frac{\gamma_{i}}{{m}_i} {{I}_i} \omega_{i},
     \label{equation 2}
 \end{equation}
 where ‘$i$’ represents individual colloids and $\gamma_{i}$ ({Supporting Information} S5) and ${m}_i$ are the friction coefficient and mass of the individual colloids respectively. The torque estimated for AT, shown in Figure \ref{fig:Figure_2}(d), is $|\tau| \simeq 1.7\cdot10^{-19}$ Nm. Although both the trimer structures in Figure \ref{fig:Figure_2}(c) and Figure \ref{fig:Figure_2}(d) have almost similar compositions, only AT in Figure \ref{fig:Figure_2}(d) exhibits rotational motion. Thus, the presence of thermally active colloid and the resultant generation of optothermal fields plays a crucial role in the rotation of the structure.

\subsection{Optothermal forces and handedness of rotation of trimer structures}

\begin{figure}[ht]
\centering
\includegraphics[width=470 pt]{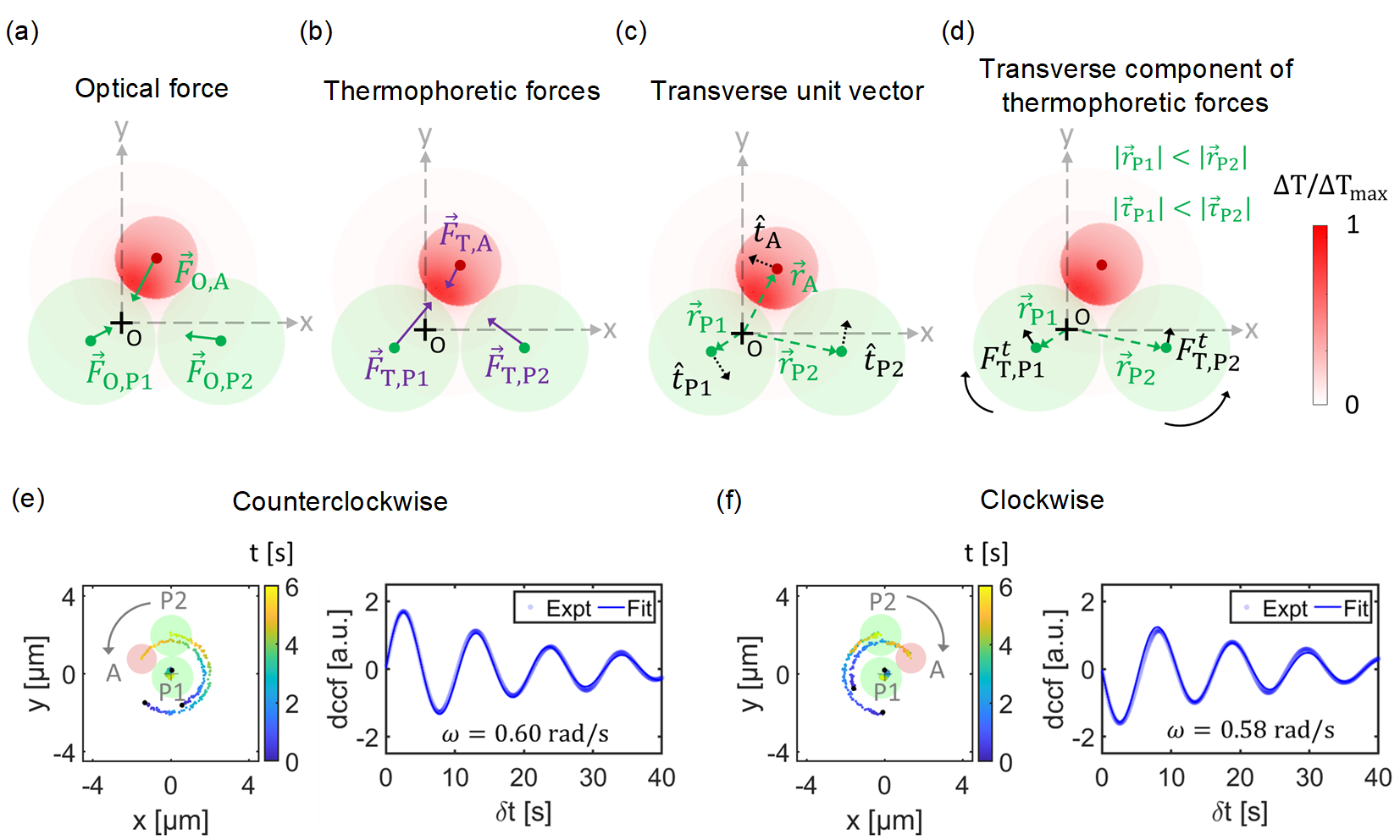}
\caption{Optothermal forces and rotation of AT. Schematic force vector diagram of (a) optical gradient forces and (b) thermophoretic forces in AT structure, where '+' indicates the beam center. (c) Transverse unit vector at the location of individual colloids in the AT structure. (d) Transverse component of thermophoretic forces on P1 and P2. In (e) and (f) schematic trajectories of two AT structures exhibiting counterclockwise and clockwise rotation along with their calculated dccfs are shown. The corresponding angular velocity of rotation at laser intensity $=5.7$ mW/\textmu$\textrm{m}^2$ is obtained as $\omega=0.60$ rad/s and $\omega=0.58$ rad/s respectively.}
\label{fig:Figure_3}
\end{figure}
To understand the origin of rotation, we have qualitatively investigated different forces acting on the constituent colloids forming AT (shown in Figure \ref{fig:Figure_2}(d)). The schematic vector diagram of different forces acting in an AT structure are shown in Figure \ref{fig:Figure_3}(a) and \ref{fig:Figure_3}(b). In the absence of peripheral colloids, P2 and A, the optical gradient force confines P1 to the beam center. But in the presence of the peripheral colloids, the optical gradient forces on P2 and A ($\Vec{F}_\textrm{O, P2}$ and $\Vec{F}_\textrm{O, A}$, indicated by green arrows in Figure \ref{fig:Figure_3}(a)), lead to off-centered confinement of P1. Since the optical gradient forces, $\Vec{F}_{\textrm{O}, i}$ on colloid $i$ ($i$ = P1, P2 and A) with position vector $\Vec{r}_i$ (measured from beam center O), are directed towards the beam center i.e. $\Vec{F}_{\textrm{O}, i} \parallel \Vec{r}_i$, it implies they cannot induce rotation as $\Vec{\tau}=\sum_{i} \Vec{r}_i\times \Vec{F}_{\textrm{O}, i} = 0$. This is evident from the absence of rotational motion in PT. Collectively $\Vec{F}_{\textrm{O}, i}$ tends to confine the colloids to the beam center and the optical gradient force on P1 can be estimated as $F_\textrm{O, P1} = k(\textrm{I}_0) \cdot {r_c} \simeq 0.36$ pN, where ${r_c}$ and ${k(\textrm{I}_0)}$ are the radius of rotation and the trap stiffness of P1 at laser intensity $\textrm{I}_0$ ({Supporting Information} S6). In addition, due to the thermophoretic behaviour of constituent colloids {\cite{piazza2008thermophoresis, wurger_thermal_transport, opto-thermophoretic_manipulation}}, the thermal field produced by colloid A exerts thermophoretic forces $\Vec{F}_{\textrm{T}, i}$ on them, indicated by violet arrows in Figure \ref{equation 3}(b). These $\Vec{{F}}_{\textrm{T}, i}$ act from the particle center towards the positive temperature gradient and are not necessarily along radial direction. In our experimental configuration, the offset position of colloid A leads to an asymmetric temperature field around it. The side of colloid A close to the beam center has a higher temperature than the other sides ({Supporting Information} S7). This asymmetric temperature profile leads to different magnitudes of thermophoretic forces on P1 and P2 ($\Vec{{F}}_{\textrm{T, P1}}$ and $\Vec{{F}}_{\textrm{T, P2}}$). Besides, this asymmetric thermal field also leads to a self-thermophoretic force $\Vec{{F}}_{\textrm{T, A}}$ on colloid A {\cite{sano_self-thermophoresis}}, due to thermo-osmotic fluid flow around it. However the exact experimental estimation of this asymmetric thermal field and thermophoretic forces are very difficult. For simplicity, we have considered that colloid A, which is at $\delta x$ distance from the beam center, has a uniform surface temperature given by ({see Supporting Information} S7), 
\begin{equation}
    {\Delta \textrm{T}}(\delta x)={\Delta \textrm{T}_0 (\textrm{I}_0 )}  e^{-\frac{\delta x^2}{w_0^2}},
    \label{equation 3}
\end{equation}
 where $w_0$ and ${\Delta \textrm{T}_0 (\textrm{I}_0)}$ represent the beam waist and the maximum surface temperature of a centrally heated colloid A at laser intensity ${\textrm{I}_0}$ respectively. ${\Delta \textrm{T}_0 (\textrm{I}_0)}$ is estimated using the nematic to isotropic phase transition of 5CB liquid crystals ({Supporting Information} S7). The approximation of uniform surface temperature of colloid A has been considered to get an approximate quantitative measure of its maximum surface temperature and the resulting thermophoretic forces and torques in such a structure. Under this approximation, the magnitude of the thermophoretic forces on P1 and P2 (which also have the contribution from slip flow) are estimated as {\cite{direct_thermophoretic_force_helden2015}}, ${F_\textrm{T, P1}}\simeq{F_\textrm{T, P2}}\simeq -\gamma \mathrm{D_{T}} {\nabla{\textrm{T}}} \simeq 0.04$ pN, where $\mathrm{D_T}$ is the thermo-diffusion constant of P1 and P2 ({Supporting Information} S8).  These ${F_\textrm{T, P1}}$ and ${F_\textrm{T, P2}}$ are acting along the colloid A (Supporting Information S9). Here the self-thermophoretic force on colloid A has not been taken care of because of the consideration of uniform surface temperature. These thermophoretic forces on colloid $i$ ($i=$ P1, P2) can be decomposed into radial (${F}^{r}_{\textrm{T}, i}$) and transverse components (${F}^{t}_{\textrm{T}, i}$) as ${F}^{r}_{\textrm{T}, i} = \Vec{F}_{\textrm{T}, i}\cdot \hat{r}_i$ and ${F}^{t}_{\textrm{T}, i} = \Vec{F}_{\textrm{T}, i}\cdot \hat{t}_i$, where $\hat{r}_i$ and $\hat{t}_i$ are the unit vectors along radial and the transverse direction respectively (see Figure 3(c)). Only the transverse components of these thermophoretic forces (${F}^{t}_{\textrm{T}, i}$, indicated by black arrows in Figure \ref{fig:Figure_3}(d)) give rise to the torque for rotation as, $\Vec{\tau} = \sum_{i}\Vec{{r}_i}\times \Vec{{F}}_{\textrm{T}, i} = \sum_{i} {r}_i \cdot {F}^{t}_{\textrm{T}, i}$, where $i$ = P1 and P2. In this way, we have obtained ${F}^{t}_\textrm{T, P1} \simeq -0.02$ pN and ${F}^{t}_\textrm{T, P2} \simeq 0.03$ pN. The negative ${F}^{t}_\textrm{T, P1}$ and positive ${F}^{t}_\textrm{T, P2}$ is because ${F}^{t}_\textrm{T, P1}$ on P1 is antiparallel to the transverse unit vector $\hat{t}_\textrm{P1}$ at the position of P1, whereas ${F}^{t}_\textrm{T, P2}$ on P2 is parallel to the transverse unit vector $\hat{t}_\textrm{P2}$ at the position of P2. While the torque ($\tau_{\textrm{P1}}\simeq -0.03\cdot10^{-19}$ Nm) induced by ${{F}}^{{t}}_{\textrm{T, P1}}$ on P1  tends to rotate the structure in a clockwise direction, the torque ($\tau_{\textrm{P2}}\simeq 0.62\cdot10^{-19}$ Nm) induced by ${{F}}^{{t}}_{\textrm{T, P2}}$ on P2  tends to rotate the structure in a counterclockwise direction. Although the magnitude of these transverse force components ${F}^{t}_\textrm{T, P1}$ and ${F}^{t}_\textrm{T, P2}$ are almost the same, due to the significant difference in the length of the force arm ($r_\textrm{P2} > r_\textrm{P1}$), the torques induced by them are significantly different in magnitude ($\tau_\textrm{P2} > \tau_\textrm{P1}$), resulting in a net torque which rotates the structure in a counterclockwise direction. Besides the rotation, it can also be noted that due to thermophoretic forces and slip flow-induced drag forces, the colloids in AT are much more tightly bound to each other than colloids in PT ({Supporting Information} S10).

 Evidently, the rotation in AT is driven by the optothermal interaction facilitated between the constituent colloidal entities due to the asymmetric positioning of colloid A. Thus, by changing the relative position of colloid A in the AT, the sense of rotation can be controlled as shown in SI video 2. Figure \ref{fig:Figure_3}(e) shows that if colloid A is positioned to the left of P2, then the AT will rotate counterclockwise. Whereas, if colloid A is positioned to the right side of P2, the AT rotates in a clockwise direction as shown in Figure \ref{fig:Figure_3}(f). We can even dynamically change the direction of rotation by switching the laser beam spot from one passive colloid to another (see SI video 3). The angular velocity of rotation ($\omega$) of the AT structures (in Figure \ref{fig:Figure_3}(e) and \ref{fig:Figure_3}(f)) were obtained by calculating the dccf as,  $\omega=0.60$ rad/s and $\omega=0.58$ rad/s respectively. The magnitude of the angular velocity of rotation ($\omega$) is similar in both these cases but it may vary slightly among different ATs’ because of non-uniform doping and size of constituent colloids.
 
 To get further insight into the emergence of rotation in AT, we have simulated thermo-osmotic fluid flow around the structure by employing a finite element method-based numerical simulation technique ({Supporting Information} S11). The fluid-colloid interfaces are set to their respective thermo-osmotic slip coefficients ($\chi$) estimated from their thermo-diffusion constant ($\mathrm{D_T}$) as $\chi = \sfrac{3}{2} \textrm{T} \textrm{D}_\textrm{T}$ (see Supporting Information S8). Additionally, colloid A is considered to serve as a heat source with uniform surface temperature governed by equation (\ref{equation 3}). The resultant thermo-osmotic slip flow velocity comes out to be $\simeq$1 {\textmu}m/s, leading to the approximate flow-induced drag force on P1 and P2 as 0.08 pN, which is comparable to the estimated thermophoretic forces (0.04 pN). The resultant slip flow-induced torque contribution is estimated as $\simeq 10^{-19}$ Nm which is of a similar order of magnitude as the experimental torque. The temperature-dependent density of the surrounding fluidic environment can also lead to convective fluid flow in the order of 1 nm/s, which can give rise to negligible torque contribution of $\simeq 10^{-24}$ Nm. Therefore, we can attribute the dominant contribution of torque to the thermo-osmotic flow field.

\subsection{Control over the speed of rotation}
\begin{figure}[ht]
\includegraphics[width = 450 pt]{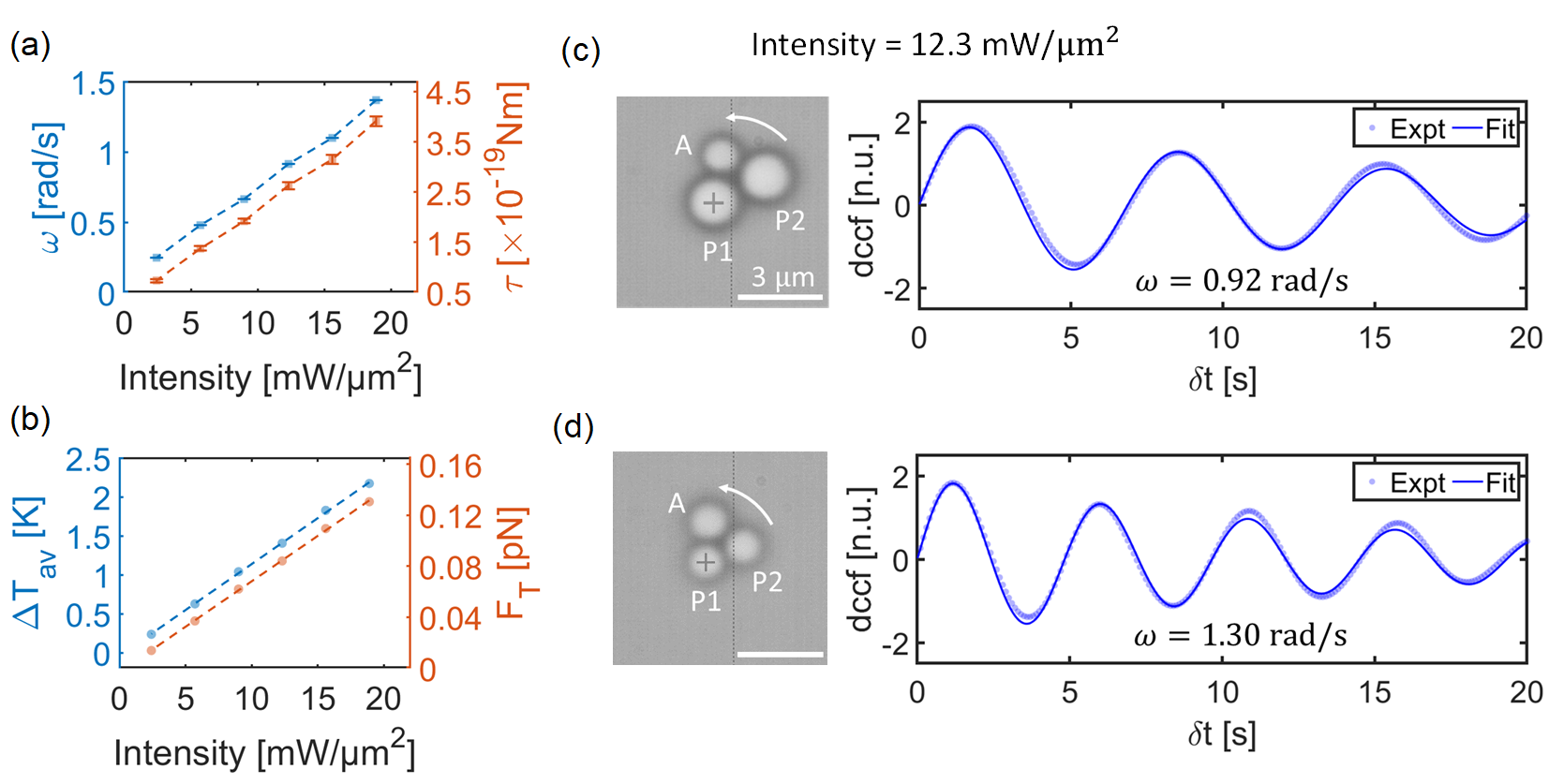}

\caption{Control over rotational dynamics of AT. The rotational dynamics of AT structure depends on the laser intensity. (a) The estimated angular velocity of rotation ($\omega$) and corresponding torque ($\tau$) show an incremental trend with laser intensity. (b) Linear increment of the average surface temperature (${\Delta \mathrm{T_{av}}}$) of the thermally active colloid (calculated from equation (\ref{equation 3})) and thermophoretic force (${F_\textrm{T}}$) with laser intensity. (c) and (d) show two AT structures, with diameter of P1 and P2 as $\mathrm{d_{P1}}=\mathrm{d_{P2}}=2.0$ {\textmu}m and $\mathrm{d_{P1}}=\mathrm{d_{P2}}=1.3$ {\textmu}m  respectively along with their corresponding calculated dccf, where the diameter of colloid A is same $\mathrm{d_{A}}=1.3$ {\textmu}m. }
\label{fig:Figure_4}
\end{figure}

 While the sense of rotation of AT can be modulated by changing the orientation of constituent colloidal entities, the intensity of the incident laser beam can be used to control the corresponding angular velocity. Figure \ref{fig:Figure_4}(a) shows that both the rotational speed as well as the torque produced in the AT increases almost linearly with increasing laser intensity. This can be attributed to the linear increase in the average surface temperature of colloid A and corresponding thermophoretic forces with the intensity of the incident laser beam (Figure \ref{fig:Figure_4}(b)). A similar linear incremental trend of the torque is also obtained from the simulation of the thermo-osmotic flow field ({Supporting Information} S11).

Besides the laser intensity, rotational dynamics of AT can also be modulated by changing the diameter of constituent passive colloids (P1, P2). This is because the thermal field, inertia, and friction coefficients of colloids in the structure get modified, as the diameter of passive colloids are changed. To see these effects, we have studied two AT structures having different sizes of passive colloids (see Figure \ref{fig:Figure_4}(c) and \ref{fig:Figure_4}(d)), where the diameter of P1 and P2 are $\mathrm{d_{P1}}=\mathrm{d_{P2}}=2.0$ {\textmu}m and $\mathrm{d_{P1}}=\mathrm{d_{P2}}=1.3$ {\textmu}m respectively (keeping diameter of colloid A same as $\mathrm{d_{A}}=1.3$ {\textmu}m). The resulting torque imbalance induced by the thermophoretic forces (${F_\textrm{T}}\simeq$ 0.08 pN and  0.13 pN) are $\tau = 1.2\cdot 10^{-19}$ Nm and $1.45\cdot 10^{-19}$ Nm for structures shown in Figure \ref{fig:Figure_4}(c) and \ref{fig:Figure_4}(d) respectively. Inserting the value of these torques into equation (\ref{equation 2}), we obtained that AT in Figure \ref{fig:Figure_4}(c) should exhibit slower rotation ($\omega \simeq 0.43$ rad/s) than AT in Figure \ref{fig:Figure_4}(d) ($\omega \simeq 1.2$ rad/s). This matches with the experimentally obtained trend of slower rotation ($\omega \simeq 0.92$ rad/s) for AT in Figure \ref{fig:Figure_4}(c) than in Figure \ref{fig:Figure_4}(d) ($\omega \simeq 1.30$ rad/s). In case of the bigger trimer, there is a significant deviation observed. This can be due to various reasons such as in our estimation, we have not considered the hydrodynamic interaction between the colloids \cite{wurger2007thermophoresis, wall_effect_on_rotating_sphere, fily2012cooperative}. During the experimental estimation of the torque using the moment of inertia ($I_i$) of individual colloids, we used equation (\ref{equation 2}). Here the friction coefficient $\gamma_i$  of individual colloids has been estimated from the free diffusion of colloids near a glass coverslip (Supporting Information S5), which can be significantly perturbed due to the presence of other colloids or surfaces nearby \cite{diffusion_correction, volpe_book}. If these considerations are taken into account, the theoretical and experimental values should converge.

\subsection{Structures with a higher number of colloids}

\begin{figure}[ht]
\includegraphics[width = 400 pt]{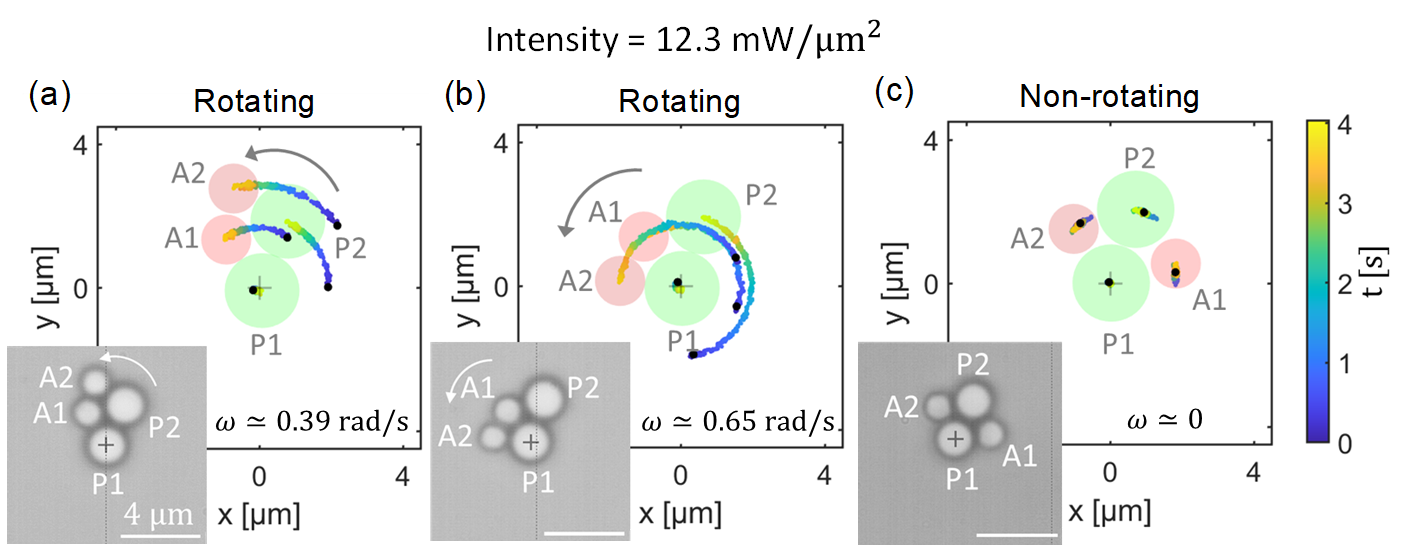}
\caption{Rotational dynamics of structures composed of a higher number of colloids. (a)-(c) Show trajectories of differently oriented structures composed of two passive MF colloids (P1, P2) and two thermally active  PS colloids (A1, A2) of diameters $\mathrm{d_{P1}}=\mathrm{d_{P2}}=2$ {\textmu}m and $\mathrm{d_{A1}}=\mathrm{d_{A2}}=1.3$ {\textmu}m respectively. Insets show bright field optical images. Structures shown in (a) and (b) exhibit rotational dynamics with an angular velocity of rotation of 0.39 rad/s and 0.65 rad/s respectively, whereas structure in (c) does not.}
\label{fig:Figure_5}
\end{figure}

Since the induced thermal field plays a key role in the rotational dynamics, we can use additional thermally active colloids to modulate the rotational dynamics of the structures. Here, we explore the structure that emerges in such cases as well as their resultant dynamics (see SI video 4). The asymmetrical structures in Figure \ref{fig:Figure_5}(a) and \ref{fig:Figure_5}(b), have two thermally active colloids (A1, A2) on the same side of P2, exhibiting characteristic rotation. Since colloid A2 in Figure \ref{fig:Figure_5}(a) is farther away from the beam center than in Figure \ref{fig:Figure_5}(b), the effective surface temperature rise and consequently induced thermo-osmotic slip flow will be lesser for structure in Figure \ref{fig:Figure_5}(a). As a result, the structure in Figure \ref{fig:Figure_5}(a) exhibits slower rotation (0.39 rad/s) than that in Figure \ref{fig:Figure_5}(b) (0.65 rad/s). In contrast, when colloids A1 and A2 are positioned on both sides of P2 as in Figure \ref{fig:Figure_5}(c), then the thermophoretically induced torques due to colloid A1 and A2 will cancel each other resulting in the absence of rotational motion in the structure. The temporal evolution of such a structure, depicted in {Supporting Information} S12, shows rotational motion until the optothermal asymmetry and hence thermophoretic force-induced torque imbalance is maintained.

\section{Conclusion}
In conclusion, we have experimentally demonstrated the emergence of directional rotational motion of multi-component microscopic colloidal structures without any direct angular momentum transfer from optical fields. We have also shown that in our system we can dynamically control the handedness of rotation of the structure by changing its optothermal asymmetry. Our approach provides a clear advantage over fabricated asymmetric structures in view of dynamic control on the handedness of rotation {\cite{single_beam_rotation_Higurashi, single_beam_rotation_Pal_Ormos, wittmeier2015rotational, wang2017silicon, microscopic_engine_schmidt2018, peng2020opto}}. Further, we have also shown how the intensity of the incident optical field can be used as an environmental cue to modulate the rotational dynamics of optothermally assembled asymmetrical structures. Our approach of emulating rotational dynamics in thermally active-passive colloidal structures can be implemented in a wide variety of colloids in view of their thermophoretic nature. Similar directional rotations are also observed in natural out-of-equilibrium systems such as bacteria {\cite{bacterial_colony_formation_self_generated_vortices, bacterial_self_concentration_RE_Goldstein, confinement_bacteria_suspension_votex}} and sperm cells {\cite{self-organized_vortex_sperm, cooperation_sperm_cell, sperm_cell_swimming_symmetry_2020}}. Thus, our approach can be a test bed for understanding the dynamics of such out-of-equilibrium systems and the influence of environmental cues on their dynamics. Our findings can potentially be useful to study the interaction between collective self-propelling organism, and out-of-equilibrium thermodynamics at the microscale.

\begin{acknowledgement}
 R.C. and C.E.R. thank Ashutosh Shukla and Sumant Panday for the fruitful discussion. G.V.P.K. also acknowledges Prof. Deepak Dhar from IISER Pune, Prof. Vijaykumar Krishnamurthy from I.C.T.S, Bengaluru, and Dr. Shradha Mishra from I.I.T.(BHU) for valuable discussion.

\end{acknowledgement}
\section{Funding sources}
This work was partially funded by AOARD (grant number FA2386-22-1-4017) and Swarnajayanti fellowship grant (DST/SJF/PSA-02/2017-18).
\begin{suppinfo}
Supporting information containing the following information is available with the manuscript. S1: Scanning electron micrograph and ultraviolet-visible spectra of thermally active PS colloid, S2: Experimental setup, S3: Experimental laser beam profile, S4: Estimation of experimental torque, S5: Diffusion constant and friction coefficient of colloids, S6: Optical trapping of colloids, S7: Thermal field around thermally active PS colloid, S8: Thermo-diffusion constant and thermo-osmotic slip coefficient of colloids, S9: Thermophoretic forces, S10: Interparticle distances in active and passive trimers, S11: Simulation of flow fields, S12: Temporal evolution of optothermally asymmetric structure.

Supporting videos: SI video 1: Dynamics of active-passive structures under focused illumination, SI video 2: Handedness of rotation of AT structures, SI video 3: Dynamic change of direction of rotation, SI video 4: Dynamics of structures with more than three colloids.

To avail the supporting information and videos please click \url{https://drive.google.com/drive/folders/1Vnb735moAL8HBPDilJ5GIcjRWdUaEaX4?usp=sharing}.

\end{suppinfo}


\providecommand{\latin}[1]{#1}
\makeatletter
\providecommand{\doi}
  {\begingroup\let\do\@makeother\dospecials
  \catcode`\{=1 \catcode`\}=2 \doi@aux}
\providecommand{\doi@aux}[1]{\endgroup\texttt{#1}}
\makeatother
\providecommand*\mcitethebibliography{\thebibliography}
\csname @ifundefined\endcsname{endmcitethebibliography}  {\let\endmcitethebibliography\endthebibliography}{}

\end{document}